\documentclass[journal=jacsat,manuscript=article]{achemso}

\usepackage{xcolor,rotating,graphicx}
\usepackage{amsmath,amssymb,ulem}
\usepackage{dsfont}
\usepackage{epsf}
\usepackage{bm}

\renewcommand{\Im}{{\rm Im}}
\newcommand{\ri}{{\rm i}}
\newcommand{\re}{{\rm e}}
\newcommand{\rd}{{\rm d}}

\newcommand{\rs}{{\rm s}}
\newcommand{\rp}{{\rm p}}

\newcommand{\kb}{k_{\rm B}}

\author{Svend-Age Biehs}
\email{s.age.biehs@uni-oldenburg.de (S.-A.Biehs)}
\affiliation{Institut f\"{u}r Physik and Center of Interface Science, Carl von Ossietzky Universit\"{a}t, 26111, Oldenburg, Germany}
\author{Igor V. Bondarev}
\email{ibondarev@nccu.edu (I.V.Bondarev)}
\affiliation{Department of Mathematics \& Physics, North Carolina Central University,\\ Durham, NC 27707, USA}

\title{Far- and Near-Field Heat Transfer in Transdimensional Plasmonic Film Systems}

\begin{document}
%%%%%%%%%%%%%%%%%%%%%%%%%%%%%%%
%% The "tocentry" environment can be used to create an entry for the graphical table of contents. It is given here as some journals require that it is printed as part of the abstract page. It will be automatically moved as appropriate.
%%%%%%%%%%%%%%%%%%%%%%%%%%%%%%%%%

\begin{abstract}
We compare the confinement-induced nonlocal electromagnetic response model to the standard local Drude model routinely used in plasmonics. Both of them are applied to study the heat transfer for transdimensional plasmonic film systems. The former provides greater Woltersdorff length in the far-field and larger film thicknesses at which heat transfer is dominated by surface plasmons, leading to enhanced near-field heat currents. Our results show that the nonlocal response model is capable of making a significant impact on the understanding of the radiative heat transfer in ultrathin films.
\end{abstract}


\newpage

\subsection{Introduction}

Present-day nanofabrication techniques make it possible to produce ultrathin films of precisely controlled thickness down to a few monolayers~\cite{thingold,Shah17,Shah18,ZhelNatCom18,MariaNL19,javierOptica19,GarciaAbajoACS19,TunableGold,SnokeBond21,NL22TiN}.~Often referred to as transdimensional (TD) quantum materials~\cite{BoltShalACS19,CNArr21PRAppl,BondADP22,BondPRR20,NL22TiN}, such films offer high tailorability of their electronic and optical properties not only by altering their chemical and electronic composition (stoichiometry, doping) but also by varying their thickness (the number of monolayers)~\cite{BondPRR20,BondOMEX17,BondMRSC18,BondOMEX19,CommPhys-bond,magnons,Manjavacas22}.

Plasmonic TD materials (ultrathin metallic films) are irreplaceable for studies of the fundamental properties of the light-matter interaction as it evolves from a single 2D atomic layer to a larger number of layers approaching the 3D bulk material properties~\cite{BoltShalACS19,BondPRR20}. They offer controlled light confinement and large tailorability of their optical properties due to their thickness-dependent localized surface plasmon (SP) modes~\cite{CNArr21PRAppl,BondADP22,BondPRR20,BondOMEX17,BondMRSC18,BondOMEX19,CommPhys-bond,magnons,Manjavacas22}. The strong vertical quantum confinement makes these modes distinct from those of conventional thin films commonly described either by 2D or by 3D material properties with boundary conditions on their top and bottom interfaces~\cite{Ritchie,Economou,Dahl,Theis,Ando,Chaplik,Wang,Pitarke,Politano}. Their properties can be understood in terms of the confinement-induced nonlocal Drude electromagnetic response theory proposed~\cite{BondOMEX17} and verified experimentally~\cite{NL22TiN,Lavrinenko19} recently. The electromagnetic response nonlocality was earlier reported experimentally to be a remarkable intrinsic property of quantum-confined metallic nanostructures~\cite{Brongersma17,Koppens18}. It is this nonlocality that enables a variety of new quantum phenomena in ultrathin TD plasmonic film systems, including the thickness-controlled plasma frequency red shift~\cite{BondOMEX17,NL22TiN}, the low-temperature plasma frequency dropoff~\cite{Lavrinenko19}, the SP mode degeneracy lifting~\cite{BondPRR20,CampionePRB15}, a series of quantum-optical~\cite{BondADP22} and nonlocal magneto-optical effects~\cite{BondMRSC18}, as well as quantum electronic transitions that are normally forbidden~\cite{Rivera,CNArr21PRAppl,CNArr21JAP}.

The confinement-induced nonlocal Drude EM response theory is built on the Keldysh-Rytova (KR) pairwise electron interaction potential~\cite{BondOMEX17} (and so referred to as the KR model in what follows for brevity). The KR interaction potential takes into account the vertical electron confinement~\cite{KRK,KRR}, which makes it much stronger than the electron-electron Coulomb potential~\cite{KRK}, offering also the film thickness as a parameter to control the nonlocal optical response of the TD plasmonic films. We perform a comparative study of the far-field and near-field heat transfer processes in the metallic TD film systems using the nonlocal KR model and the standard local Drude EM response model (a 'workhorse' routinely used in plasmonics). We show that the nonlocal KR model results in the greater Woltersdorff length (the film thickness at which its thermal emission is maximal) in the far-field regime and predicts larger film thicknesses at which the near-field heat transfer starts being dominated by surface plasmons, as compared to those resulted from the local Drude model. This can lead to an enhanced near-field heat flux (HF) between two ultrathin metallic films or materials with ultrathin metal coating. Our study thereby suggests that the theoretical treatment and experimental data interpretation for the radiative heat transfer processes in planar TD plasmonic nanostructures is to include the nonlocal effect as described by the KR model in order to provide reliable results.

\section{Confinement-induced nonlocality}

%%%%%%%%%%%%%%%%%%%%%%%%%%%%%%%%%%%%%%%%%%%%%%%%%%%%%%%%%%%%%%%%%%%%%%%%%%%%%%%%%%
%%% Keldysh Potential
%%%%%%%%%%%%%%%%%%%%%%%%%%%%%%%%%%%%%%%%%%%%%%%%%%%%%%%%%%%%%%%%%%%%%%%%%%%%%%%%%%

\begin{figure}[t]
	\center
	\includegraphics[width=0.75\textwidth]{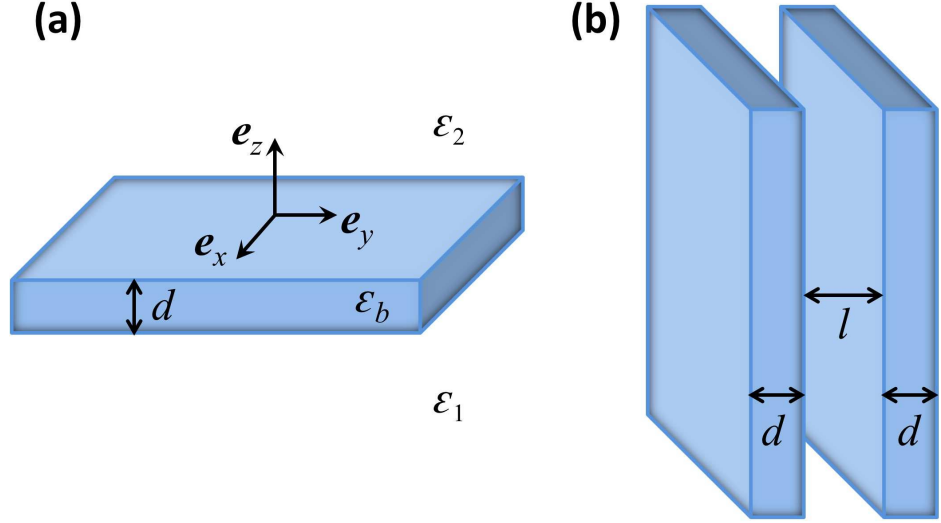}
	 \caption{The geometries of the problems studied for far-field (a) and near-field (b) thermal radiation transfer in the TD plasmonic film systems.}
	\label{fig1}
\end{figure}

If the environment has a lower dielectric constant than that of the film, such as the case of the finite-thickness metallic slab shown in Fig.~\ref{fig1}~(a), where $\varepsilon_{1,2}\!<\!\varepsilon_b$ are, respectively, the static permittivities of the substrate, the superstrate and the slab (contributed by the positive ion background and interband electronic transitions), then the increased 'outside' contribution makes the Coulomb interaction of electrons vertically confined inside the slab stronger than that in a homogeneous medium with the same permittivity~\cite{KRK}. The interaction potential takes the KR form to give the in-plane plasma oscillation frequency as follows~\cite{BondOMEX17}
\begin{equation}
\omega_p=\omega_p(k)=\frac{\omega_p^{3D}}{\sqrt{1+1/\tilde{\varepsilon}kd}}\,.
\label{omegapkd}
\end{equation}
Here, $\tilde{\varepsilon}\!=\!\varepsilon_b/(\varepsilon_1+\varepsilon_2)$ is the relative dielectric constant of the film, $d$ is its thickness, and $k$ is the in-plane momentum absolute value. With this plasma frequency, the low-energy (well below the first interband transition energy) in-plane EM response function of the plasmonic film takes the form
\begin{equation}
\frac{\epsilon(k,\omega)}{\varepsilon_b}=1-\frac{\omega_p^{2}(k)}{\omega(\omega+i\Gamma_D)},
\label{Lindhard}
\end{equation}
which is $k$-dependent and so spatially dispersive, or nonlocal. Here, the damping constant $\Gamma_D$ stands to (phenomenologically) include the  inelastic electron scattering rate. It can be seen that as $d$ decreases, $\omega_p(k)$ shifts to the red and Eq.~(\ref{omegapkd}) acquires the $\sqrt{k}$-type nonlocal spatial dispersion of 2D materials. In the TD regime, for ultrathin plasmonic films of finite thickness, the low-frequency response is controlled by Eqs.~(\ref{omegapkd}) and (\ref{Lindhard}), with $\varepsilon_b\!\sim\!10$ contributed by both positive ion background and interband transitions~\cite{NL22TiN,Lavrinenko19}. As $d$ increases and gets large, Eq.~(\ref{omegapkd}) can be seen to gradually approach $\omega_p^{3D}$, the bulk material screened plasma frequency, and Eq.~(\ref{Lindhard}) takes the local Drude form. This is the essence of the nonlocal KR model of the plasmonic TD film EM response.

The degree of the confinement-induced nonlocality associated with $k$-dependence in Eq.~(\ref{Lindhard}) can be evaluated by effecting the inverse Fourier transformation from the reciprocal 2D space to the direct coordinate 2D space. This gives the coordinate/frequency-dependent specific permittivity (per unit area) in the form~\cite{BondMRSC18}
\begin{equation}
\frac{\varepsilon(\bm{\rho},\omega)}{\varepsilon_b}=\delta(\rho)\!-\frac{\big(\omega_p^{3D}\big)^2}{2\pi\omega(\omega+i\Gamma_D)}\int_{k_m}^{k_c}\!\!\!dk\,\frac{kJ_0(k\rho)}{1+1/\tilde{\varepsilon}kd}\,,
\label{Lindrho}
\end{equation}
where $k_c$ and $k_m(=\!2\pi/L$ with $L$ being the film lateral size) stand for the plasmon upper and lower cut-off wave vectors, respectively. The latter indicates that the largest possible plasma wavelength cannot exceed the lateral size of the film. For thick films with $\tilde{\varepsilon}kd\gg\!1$, Eq.~(\ref{Lindrho}) gives the local response of the form
\[
\frac{\varepsilon(\bm{\rho},\omega)}{\varepsilon_b}=\left[1-\frac{\big(\omega_p^{3D}\big)^2}{\omega(\omega+i\Gamma_D)}\right]\delta(\rho)
\]
For ultrathin films with $\tilde{\varepsilon}kd\ll\!1$, one obtains ($L\!\rightarrow\!\infty$)
\[
\frac{\varepsilon(\bm{\rho},\omega)}{\varepsilon_b}=1-\frac{\tilde{\varepsilon}k_cd\big(\omega_p^{3D}\big)^2}{\omega(\omega+i\Gamma_D)}\left[\delta(\bm{\rho})\!-\frac{\cos\big(k_c\rho+3\pi/4\big)}{\sqrt{2\pi^3k_c}\,\rho^{5/2}}\right]
\]
under extra condition $k_c\rho\!>\!1$ to specify the distances at which the electrostatic repulsion of electrons is screened and their motion in the form of collective plasma oscillations is well defined. The ultrathin film permittivity can now be seen to decay with distance as $\rho^{-5/2}$ due to the confinement-induced EM response nonlocality of the KR model we use.

The damping constant $\Gamma_D$ for the ultrathin films can be expressed as~\cite{NL22TiN}
\begin{equation}
  \Gamma_D = \Gamma_{\rm bulk} + \frac{\hbar v_{\rm F}}{l_{\rm eff}} + \frac{A}{d}
\label{GammaD}
\end{equation}
with the bulk damping $\Gamma_{\rm bulk}$, the surface roughness correction $A$, and the
surface scattering determined by the effective mean free path
\begin{equation}
  l_{\rm eff} = d \biggl[ 1 + \ln\biggl( \frac{l_{\rm bulk}}{d}\biggr) \biggr]
\end{equation}
with the bulk mean free path $l_{\rm bulk}$ and the Fermi velocity
\begin{equation}
  v_{\rm F} = \frac{\overline{\omega}_p}{k_c}
\end{equation}
introducing the thermally averaged plasma frequency of Eq.~(\ref{omegapkd}) in the TD regime~\cite{BondPRR20,NL22TiN}
\begin{equation}
  \overline{\omega}_p  = \frac{2 C^2 d^2 \omega_p^{\rm 3D}}{(1 + 2 Cd)\sqrt{Cd(1 + Cd)} + \sinh^{-1}(\sqrt{Cd})}
\end{equation}
with $C = \tilde{\varepsilon}k_c$. The local Drude model can be retrieved from the above expressions in the limit $d\!\rightarrow\!\infty$, which leads to $\omega_p^{\rm 3D}$ for both $\omega_p(k)$ and $\overline{\omega}_p$.

In our comparative study below we use TiN, Au, and Ag parameters taken from experimental data and listed in the Table for convenience~\cite{NL22TiN,OlmonEtAl2012,YangEtAl2015}. Transition metal nitrides have emerged as a new class of materials with great promise to substitute noble metals such as gold and silver~\cite{BoltassevaAtwater}, which have exceptional plasmonic properties but relatively low melting temperatures making them incompatible with semiconductor fabrication technologies. Titanium nitride has low-loss plasmonic response and high melting point, making it structurally stable TD material capable of forming stochiometrically perfect ultrathin films of controlled thickness down to one nanometer at room temperature. For the sake of convenience as surface roughness correction $A$ we use throughout the manuscript the value $A=0.12\;{\rm eV}\;{\rm nm}$ as experimentally found for nanometer-thick TD TiN films in Ref.~\cite{NL22TiN}. This value of the surface roughness is of course highly dependent on the specific sample, but also the Drude parameters show some variance for different samples~\cite{OlmonEtAl2012, YangEtAl2015} and in particular for thin films~\cite{LangEtAl2013}.

\begin{table}[t]
\begin{tabular}{|c|c|c|c|c|c|c|}
\hline
     & $v_{\rm F}$ & $\epsilon_b$ & ~$\omega_p^{\rm 3D}$ & $k_c$ & ~$l_{\rm bulk}$~ & ~$\Gamma_{\rm Bulk}$~ \\
     & ($\times 10^6$m/s) &       &  ~(eV)~           & ~(nm$^{-1}$) & (nm) &  (eV) \\ \hline
  Au  & 1.39 & ~9.5 & 2.9 & 3.3 & 42 & 0.02\\ \hline
  Ag  & 1.39 & ~4.2 & 4.5 & 5.2 & 56 & ~0.016\\ \hline
  TiN & 0.67 & ~9.1 & 2.5 & 6.0 & 45 & 0.2 \\ \hline
\end{tabular}
\label{Tab:parameters}
\end{table}

%%%%%%%%%%%%%%%%%%%%%%%%%%%%%%%%%%%%%%%%%%%%%%%%%%%%%%%%%%%%%%%%%%%%%%%%%%%%%%%%%%
%%% Far-field thermal radiation
%%%%%%%%%%%%%%%%%%%%%%%%%%%%%%%%%%%%%%%%%%%%%%%%%%%%%%%%%%%%%%%%%%%%%%%%%%%%%%%%%%

\section{Far-field thermal radiation}

We consider a thin film as used in transmission electron microscopy or soft X-ray spectral purity filter applications~\cite{VanZwolEtAl2015}; see Fig.~\ref{fig1}(a). The HF $\Phi_{\rm sf}$ of a single film at temperature $T_{\rm f}$, radiating in its surrounding at temperature $T_{\rm b}$, is given by~\cite{SAB2007}
\begin{equation}
  \Phi_{\rm sf}  = \int_0^\infty \frac{\rd \omega}{2 \pi} \bigl[ \Theta(\omega,T_{\rm f}) - \Theta(\omega,T_{\rm b}) \bigr]
		  \int_0^{k_0}\frac{\rd k}{2 \pi}\, k \sum_{i = \rs, \rp} e_i
\label{phisf}
\end{equation}
with $k$ and $k_0 = \omega/c$ being the in-plane and vacuum wave vector absolute values, respectively. The thermal emission is driven by the temperature difference of a harmonic oscillator
\begin{equation}
   \Theta(\omega,T) = \frac{\hbar \omega}{\re^{\hbar \omega / \kb T} - 1}.
\end{equation}
For a given polarization $i = \rs,\rp$, the film emissivity $e_i(\omega,k)=1 - |R_i|^2 - |T_i|^2$, where $R_i$ and $T_i$ are the Fresnel reflection and transmission coefficients at a given frequency $\omega$ and the incidence angle (represented here by $k$). They are given by
\begin{align}
  R_i &= \frac{r_i \bigl(1 - \re^{2 \ri k_z d} \bigr)}{1 - r_i^2  \re^{2 \ri k_z d}}, \\[0.25cm]
  T_i &= \frac{(1 - r_i^2) \re^{\ri k_z d}}{1 - r_i^2  \re^{2 \ri k_z d}},
\end{align}
where $k_z = \sqrt{k_0^2\,\epsilon(\omega,k) - k^2}$ and $r_i$ are the interface Fresnel reflection coefficients defined as
\begin{align}
  r_\rs &= \frac{k_v - k_z}{k_v + k_z}, \\[0.25cm]
  r_\rp &= \frac{k_v \epsilon(\omega,k)- k_z }{k_v \epsilon(\omega,k) + k_z }
\end{align}
with $k_v = \sqrt{k_0^2 - k^2}$. If, for example, $|R_{i=s,p}|^2 = |T_{i=s,p}|^2 = 0$ for all $\omega$ and $k$, then
\begin{equation}
   \Phi_{\rm BB} = \sigma (T_{\rm f}^4 - T_{\rm b}^4)
\end{equation}
with $\sigma = 5.6705\times10^{-8} {\rm W} \, {\rm m}^{-2} \, {\rm K}^{-4}$ being the Stefan-Boltzmann constant, to give the blackbody HF $\Phi_{\rm BB}=64.4\,{\rm W}/{\rm m}^2$ for $T_{\rm f} = 310\,{\rm K}$ and $T_{\rm b} = 300\,{\rm K}$.

\begin{figure}[t]
	\center
	\includegraphics[width=0.75\textwidth]{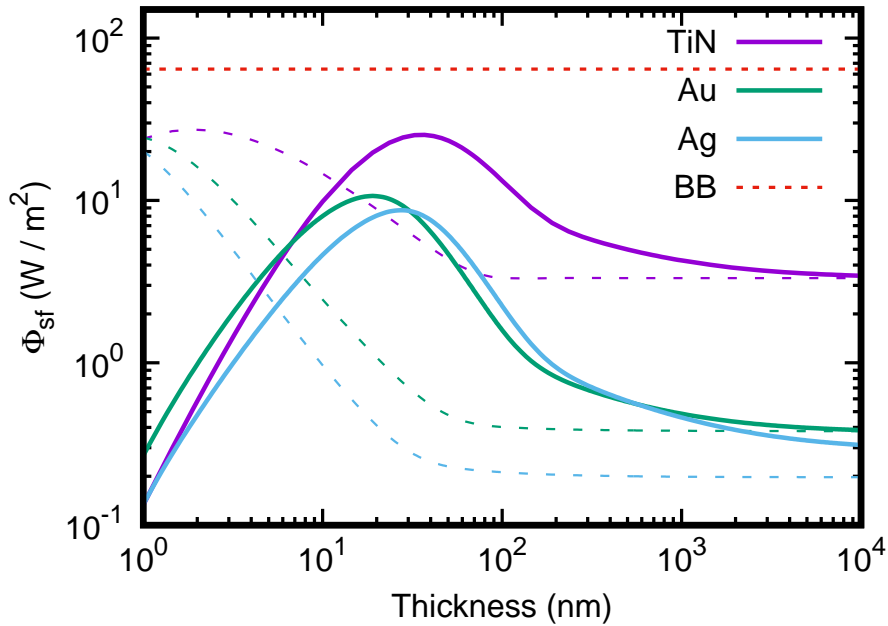}
	 \caption{The behavior of $\Phi_{\rm sf}$ for the local (dashed lines) and nonlocal (solid lines) case of a free standing TiN, Au, Ag layer as function of the film thickness $d$ with $T_{\rm f} = 310\,{\rm K}$ and $T_{\rm b} = 300\,{\rm K}$.}
	\label{fig2}
\end{figure}

Figure~\ref{fig2} compares the film thickness dependence of the HF from Eq.~(\ref{phisf}) for TiN, Au, and Ag films using the local Drude model and the non-local KR model. For large $d$ the HF converges to a constant bulk sample value. This value highly depends on the damping, the bulk damping $\Gamma_{\rm bulk}$ in this case, which determines the absorptivity and therewith the emissivity, so that the higher is $\Gamma_{\rm bulk}$ the stronger is the thermal emission. Nonetheless, in all three cases considered the radiative HF is one to three orders of magnitude less than the blackbody value, confirming that metals are poor thermal emitters due to their high reflectivity. With film thickness reduction the HF starts increasing, reaches a maximum, and then declines. This effect is known for a long time~\cite{Woltersdorff1934}. It was discussed within the framework of fluctuational electrodynamics in Ref.~\cite{SAB2007} and was measured for Ru and Pt thin films previously (see Refs.~\cite{VanZwolEtAl2015,MahanMarple1983}). The maximum is known to take place at a length scale called the Woltersdorff length $d_{\rm max}$ which for metals is typically very small and within the local Drude model in the limit $\omega \ll \Gamma_{\rm bulk} \ll \omega_p$ can be written as~\cite{SAB2007}
\begin{equation}
  d_{\rm max} \approx \frac{2 c \Gamma_{\rm bulk}}{\omega_p^2}.
\label{Eq:Woltersdorfflength}
\end{equation}

It can be seen in Fig.~\ref{fig2} that $d_{\rm max}$ of the local model is far below 10~nm for all three materials, whereas in the non-local model it shifts towards thicknesses larger than 10~nm. For TiN, in particular, it shifts from $d_{\rm max} = 1.8\,{\rm nm}$ (local model) to $d_{\rm max} = 38\,{\rm nm}$ (non-local model), whereby the nonlocal KR model predicts over an order of magnitude Woltersdorff length increase as compared to the conventional local Drude model. This can be understood from Eq.~(\ref{Eq:Woltersdorfflength}), where $\omega_p \propto \omega_p^{3D}\sqrt{k d}$ with $k d \ll 1$ and the damping is dominated by $\Gamma_D \propto A/d$ which is the case for the ultrathin TD films described by the non-local KR model as given by Eqs.~(\ref{omegapkd}) and (\ref{GammaD}), respectively. Clearly, $d_{\rm max}$ has the tendency to increase for thinner films then. As a consequence, for $d > 10\,{\rm nm}$ larger HF can be found in the non-local than in local model. For $20\,{\rm nm}$ thick Au and Ag films the HF predicted by the KR model is one order of magnitude greater than that of the local model. As the Woltersdorff length is $\sim\!10\div100$~nm in the non-local model, the predicted HF enhancement effect might be used to test this model experimentally with well-characterized ultrathin film samples. Note that the KR model is valid for $k = 2 \pi/\lambda \geq 2 \pi/L$, and so our results here hold for samples with $L \geq \lambda_{\rm th}$, where $\lambda_{\rm th}$ is the dominant thermal wavelength determined by the Wien's law.

%%%%%%%%%%%%%%%%%%%%%%%%%%%%%%%%%%%%%%%%%%%%%%%%%%%%%%%%%%%%%%%%%%%%%%%%%%%%%%%%%%
%%% Near-field thermal radiation
%%%%%%%%%%%%%%%%%%%%%%%%%%%%%%%%%%%%%%%%%%%%%%%%%%%%%%%%%%%%%%%%%%%%%%%%%%%%%%%%%%

\section{Near-field thermal radiation}

Here we aim to study the radiative heat exchange between the two thin films separated by distances smaller than the room-temperature thermal wavelength $\lambda_{\rm th} \approx 10\,\mu{\rm m}$; see Fig.~\ref{fig1}(b). In this near-field regime, as opposed to the far-field regime discussed above, the blackbody HF value is no longer a limit due to the dominant contribution of evanescent waves unaccounted for by the Stefan-Boltzmann law~\cite{PvH1971}. The near-field thermal radiation enhancement was previously studied in a great number of experiments measured the HF between two planar samples~\cite{HuEtAl2008,OttensEtAl2011, Kralik2012, LimEtAl2015, WatjenEtAl2016,BernadiEtAl2016,SongEtAl2016,Kralik2017, FiorinoEtAl2018a}. For thin films supporting surface phonon polaritons, an enhancement factor of 700 times over the blackbody value was found for separations under 30~nm~\cite{FiorinoEtAl2018b}. For metals there is also a near-field enhancement due to the $\rs$-polarized (magnetic) contribution of eddy currents~\cite{Volokitin2007,ChapuisEtAl2008}, first reported by Polder and van Hove~\cite{PvH1971}, in contrast to the $\rp$-polarized (electric) surface phonon-polariton or SP contributions widely discussed for thin films~\cite{Volokitin2007,SAB2007,SAB2007b}. For ultrathin TD metallic films the low-frequency SP contribution can also show an enhancement effect~\cite{BondPRR20,SAB2007,SAB2007b}. Similar effects for phonon-polaritonic films are predicted~\cite{FrancoeurEtAl2008,PBA2009} and experimentally verified~\cite{SongEtAl2015} as well.

\begin{figure}[t]
	\center
	\includegraphics[width=0.75\textwidth]{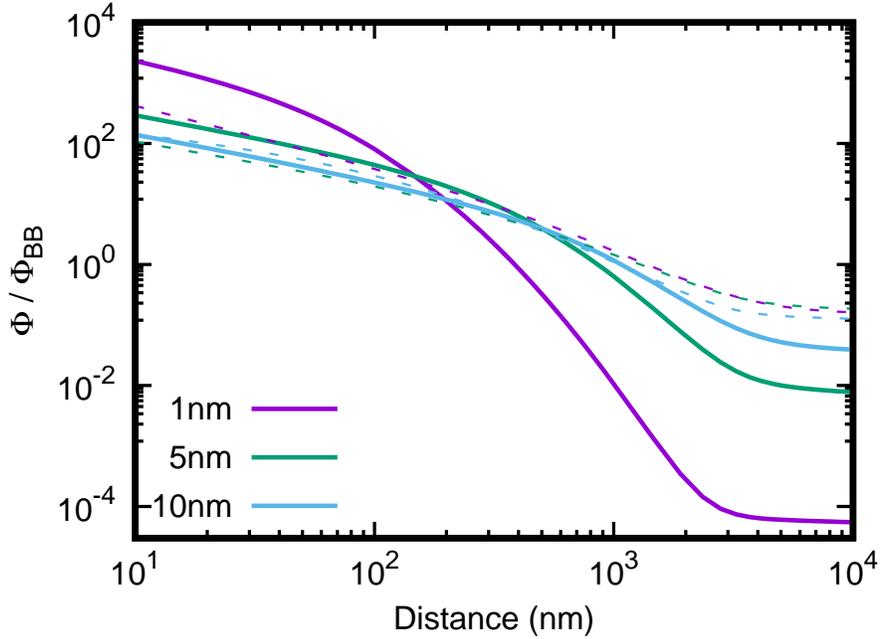}
	 \caption{Heat flux enhancements $\Phi_{\rm nonloc}/\Phi_{\rm BB}$ (solid lines) and $\Phi_{\rm loc}/\Phi_{\rm BB}$ (dashed lines) between two free standing same-thickness TiN films as functions of the interfilm separation $l$ for different film thicknesses $d = 1\,{\rm nm}$, 5nm, 10nm at $T_{1} = 310\,{\rm K}$ and $T_{2} = 300\,{\rm K}$.}
	\label{fig3}
\end{figure}

The HF between the two identical thin films at temperatures $T_1$ and $T_2$ is given by~\cite{PvH1971,PBA2009}
\begin{equation}
  \Phi = \int_0^\infty \frac{\rd \omega}{2 \pi} \bigl[ \Theta(\omega,T_{1}) - \Theta(\omega,T_{2}) \bigr]
        \int_0^\infty \frac{\rd k}{2\pi}\,k \sum_{i = \rs, \rp} \mathcal{T}_i
\label{Eq:Phi}
\end{equation}
with transmission coefficients of different form for the propagating ($k < k_0$) and evanescent ($k > k_0$) waves as follows ($i = \rs,\rp$)
\begin{equation}
  \mathcal{T}_i = \begin{cases}
                      \frac{\displaystyle\bigl(1 - |R_i|^2 - |T_i|^2\bigr)^2}{\displaystyle|1 - R_i^2 \re^{2 \ri k_v l}|^2}, & k < k_0 \\[0.25cm]
                      \frac{\displaystyle4\Im(R_i)^2 \re^{-2 k_v'' l}}{\displaystyle|1 - R_i^2 \re^{2 \ri k_v l}|^2}, & k > k_0
                   \end{cases}
 \label{Eq:TransmissionNF}
\end{equation}
Figure~\ref{fig3} compares $\Phi_{\rm loc}/\Phi_{\rm BB}$ and $\Phi_{\rm nonloc}/\Phi_{\rm BB}$ computed from these equations for two TiN films of fixed $d$ within the local and non-local model, respectively, as functions of the interfilm distance $l$. The near-field enhancement $\Phi_{\rm nonloc}/\Phi_{\rm BB}$ can be seen very high for TD film systems with $l\lesssim10$~nm but is extremely reduced in the far-field for $l \geq 10\, \mu{\rm m}$ in agreement with Fig.~\ref{fig2}. For ultrathin TD films separated by distances $l \leq 100\,{\rm nm}$ this effect should be easily accessible for measurements with today's experimental methods. Since $k \geq 2\pi/L$ in the non-local KR model, $\exp(-2 k_v'' l) \approx \exp(-2 k l)$ in the near-field quasi-static regime ($k \gg k_0$) in $\mathcal{T}_i$ in Eq.~(\ref{Eq:TransmissionNF}), whereby the $k$-integrand in Eq.~(\ref{Eq:Phi}) is dominated by $k \sim 1/l$. Hence, $k \geq 2\pi/L$ translates into $L \geq 2\pi l$ which is typically the case in near-field experiments. In general, for $l<l_{\rm bulk}$ the Lindhard-Mermin nonlocality might play a role as well~\cite{Volokitin2007,ChapuisEtAl2008}. In the most important low-frequency domain, where $k\le k_c$ and the interelectron interaction is screened with collective plasma oscillations well defined, this is $k^2$-infinitesimal order nonlocality similar to that due to the pressure term of hydrodynamical Drude models, which is much smaller than the confinement-induced $k$-infinitesimal order nonlocality of the KR model we discuss~\cite{BondOMEX17}. A full treatment of such nonlocalities is out of the scope of the present work.

\begin{figure}[t]
	\center
        \includegraphics[width=1.0\textwidth]{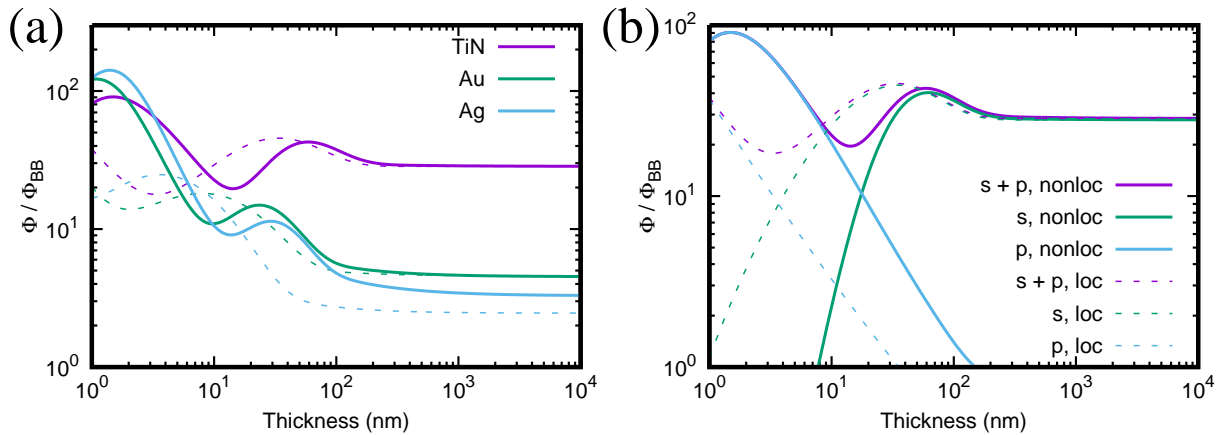}
        \caption{(a): Heat flux enhancements $\Phi_{\rm nonloc}/\Phi_{\rm BB}$ (solid lines) and $\Phi_{\rm loc}/\Phi_{\rm BB}$ (dashed lines) as functions of $d$ for pairs of same-thickness TiN, Au, and Ag films separated by $l = 100\,{\rm nm}$ at $T_{1} = 310\,{\rm K}$ and $T_{2} = 300\,{\rm K}$. (b) Contributions of the s-polarized and p-polarized waves to $\Phi_{\rm nonloc}/\Phi_{\rm BB}$ (solid lines) and $\Phi_{\rm loc}/\Phi_{\rm BB}$ (dashed lines) for TiN films in (a).}
	\label{fig4}
\end{figure}

Figure~\ref{fig4}(a) compares $\Phi_{\rm loc}/\Phi_{\rm BB}$ and $\Phi_{\rm nonloc}/\Phi_{\rm BB}$ as functions of $d$ for pairs of same-thickness TiN, Au, and Ag films separated by $l = 100\,{\rm nm}$. Depending on $d$ the KR model gives either larger or smaller HFs than the local model. For ultrathin TD films of $d < 10\,{\rm nm}$, in particular, the nonlocal model gives a much larger near-field enhancement than the local model. This enhancement is particularly strong for Au and Ag films. For the local model a similar enhancement would occur at distances smaller than $1$~nm where the macroscopic approach is no longer valid.

The HF behavior in Fig.~\ref{fig4}(a) can be explained by the transition from a thick-film regime where the s-polarization is dominant to the TD regime where the p-polarized evanescent waves prevail~\cite{BondPRR20,SAB2007,SAB2007b}. This is shown in Fig.~\ref{fig4}(b) separately for s- and p-polarized contributions in a pair of same-thickness TiN films. With $d$ decreasing, the s-polarized wave contribution first dominates the HF but then drops down abruptly, making the HF decline. The sharply increasing p-polarized wave contribution of SPs picks up and starts playing the dominant role instead. This makes the HF increase again, with much greater enhancement predicted by the nonlocal KR model than that of the local Drude model for the ultrathin TD films.

\begin{figure}[t]
	\center
	\includegraphics[width=1.0\textwidth]{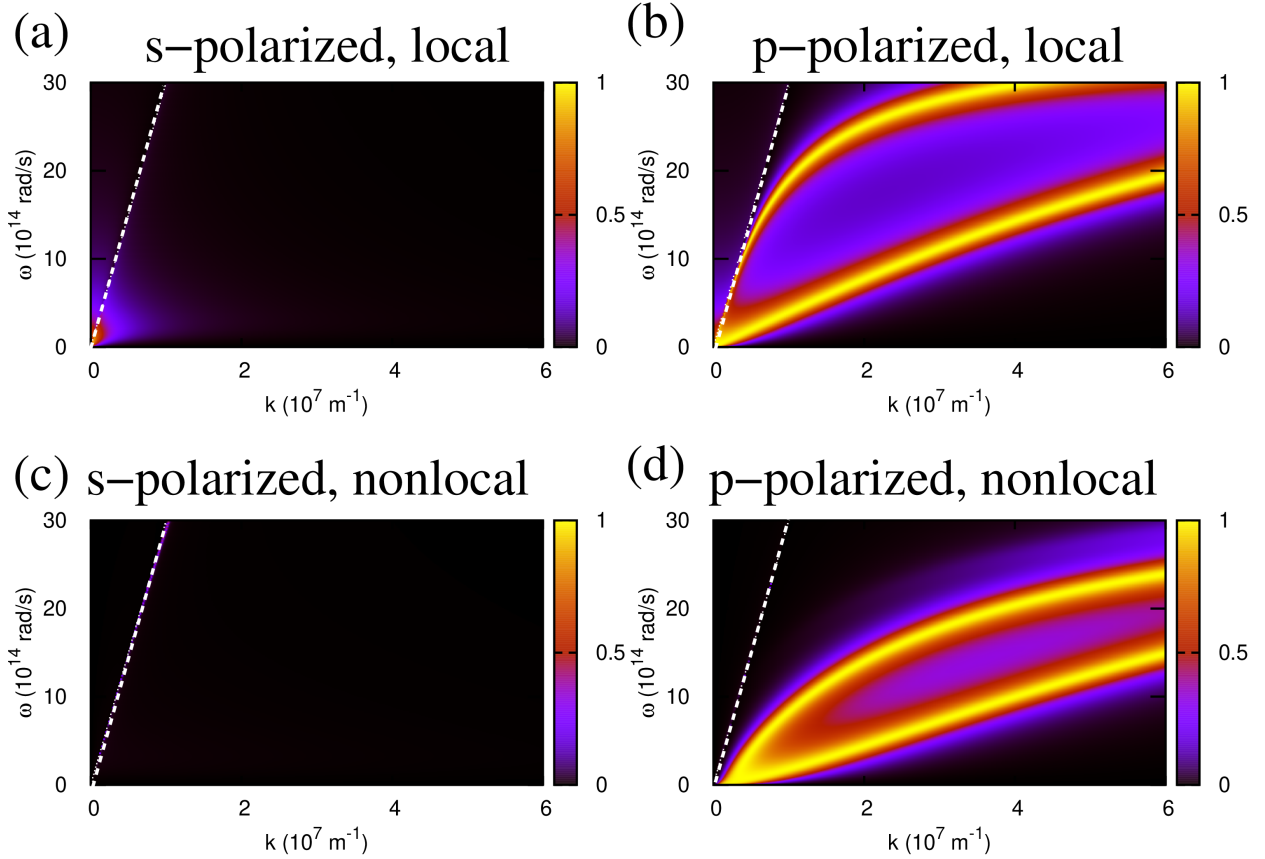}
	 \caption{Transmission coefficients $\mathcal{T}_\rs$ and $\mathcal{T}_\rp$ computed in the local Drude model (a,b) and in the nonlocal KR model (c,d) for a pair of ultrathin TiN films with $d = l = 5\,{\rm nm}$. The dashed white line is the light cone line in vacuum. The two modes seen in $\mathcal{T}_\rp$ are the long- and short-range SPs of the ultrathin TD film structures~\cite{BondPRR20,Volokitin2007}.}
	\label{fig5}
\end{figure}

The dominance of the SPs in the HF of ultrathin TD films can be directly observed in the transmission coefficients $\mathcal{T}_i$ ($i = \rs, \rp)$ as well. Figure~\ref{fig5} shows $\mathcal{T}_i$ ($i = \rs, \rp)$ in the $\omega$-$k$ space for propagating (s) and evanescent (p) waves in a pair of TiN films with $d = l = 5\,{\rm nm}$. In the local model $\mathcal{T}_\rs$ has a significant contribution from the s-polarized eddy currents, whereas for the nonlocal case this contribution does not exist for such thin layers. Contrary to the local model, in the KR model the SP modes of $\mathcal{T}_\rp$ come out red-shifted~\cite{NL22TiN,BondOMEX17} toward the thermally accessible frequency range $\sim\!2\times10^{14}\,{\rm rad/s}$ ($\lambda_{\rm th} \approx 10\,\mu{\rm m}$), explaining the greater HF enhancement predictions by the nonlocal KR model compared to the local Drude model.

\section{Conclusion}

In this Letter, we show that in the far-field regime the nonlocal KR electromagnetic response model predicts a significant increase for film thickness values at which the thermal emission of thin metallic films has a maximum, i.e.\ the Woltersdorff length is increased compared to the local Drude model. In the near-field regime the thickness of the metallic films at which the near-field heat transfer becomes SP dominated is also shifted to larger thickness values when using the nonlocal KR model. This leads to strongly enhanced near-field HFs between two thin metallic films or materials with thin metal coatings. Thus, the nonlocal KR model makes a significant impact on the understanding of thermal radiation transfer in TD materials both in the near-field and in the far-field regime. The non-local KR model can be tested in far- or near-field measurements with existing setups and well-characterized plasmonic films. Our analysis suggests that the theoretical treatment and experimental data interpretation for thin and ultrathin metallic films must incorporate the nonlocal effect as described by the KR model in order to provide reliable results in radiative heat transfer studies. Finally, the fact that the enhanced far- and near-field emission is achievable in the nonlocal KR model with much thicker films than the standard local Drude model erroneously predicts is crucial for thermal management applications with thin metallic films and coatings.

\begin{acknowledgement}
S.-A.\ B.\ acknowledges support from Heisenberg Programme of the Deutsche Forschungsgemeinschaft (DFG, German Research Foundation), project No.\ 404073166. I.V.B. is supported by the U.S. National Science Foundation under Condensed Matter Theory Program Award No.~DMR-1830874. Both authors gratefully acknowledge support from the Kavli Institute for Theoretical Physics (KITP), UC Santa Barbara, under U.S. National Science Foundation Grant No.~PHY-1748958, where this collaborative work was started. I.V.B. acknowledges KITP hospitality during his invited visit as a KITP Fellow 2022--23.
\end{acknowledgement}

\end{document}